\documentclass[11pt,a4paper]{article}
\usepackage{jheppub}
\usepackage[T1]{fontenc}
\usepackage{amsmath}
\usepackage{amssymb}
\usepackage{esint}
\usepackage{amsfonts}
\usepackage{graphicx,color}
\usepackage{slashed}
\usepackage{young}
\usepackage[vcentermath]{youngtab}
\usepackage[utf8]{inputenc}
\usepackage{tensor}
\usepackage{etoolbox}
\usepackage{bbm}
\usepackage{soul}

\makeatletter

\newcommand{\be}{\begin{equation}}
\newcommand{\ee}{\end{equation}}
\newcommand{\bea}{\begin{eqnarray}}
\newcommand{\eea}{\end{eqnarray}}

\renewcommand{\tilde}{\widetilde}
\renewcommand{\hat}{\widehat}
\newtheorem{prop}{Proposition}[section]

\newtheorem{theorem}[prop]{Theorem}

\renewcommand{\d}{\partial}

\def\cH{\mathcal{H}}

\def\cL{\mathcal{L}}

\newcommand*\xbar[1]{%
  \hbox{%
    \vbox{%
      \hrule height 0.5pt 
      \kern0.3ex
      \hbox{%
        \kern-0.0em
        \ensuremath{#1}%
        \kern-0.0em
      }%
    }%
  }%
} 

\pdfpageheight\paperheight
\pdfpagewidth\paperwidth

\pdfoutput=1 

    \patchcmd{\maketitle}{\@fpheader}{}{}{}

\title{BMS Group at Spatial Infinity: the Hamiltonian (ADM) approach}
\author[a,b]{Marc Henneaux,}
\author[c]{and C\'edricTroessaert}
\affiliation[a]{Universit\'e Libre de Bruxelles and International Solvay Institutes, ULB-Campus Plaine CP231, B-1050 Brussels, Belgium}
\affiliation[b]{Coll\`ege de France, 11 place Marcelin Berthelot, 75005 Paris, France}
\affiliation[c]{Max-Planck-Institut f\"{u}r Gravitationsphysik (Albert-Einstein-Institut),
Am M\"{u}hlenberg 1, \\ DE-14476 Potsdam, Germany}

\abstract
{New boundary conditions for asymptotically flat spacetimes  are given at spatial infinity. These boundary conditions are invariant under the BMS group, which acts non trivially. The boundary conditions fulfill all standard consistency requirements: (i) they make the symplectic form finite; (ii) they contain the Schwarzchild solution, the Kerr solution and their Poincar\'e transforms, (iii) they make the Hamiltonian generators of the asymptotic symmetries integrable and well-defined (finite).  The boundary conditions differ from the ones given earlier in the literature in the choice of the parity conditions.  It is this different choice of parity conditions that makes the action of the BMS group non trivial. Our approach is purely Hamiltonian and off-shell throughout.}

\makeatother

\begin{document}
\maketitle \flushbottom

\section{Introduction}
\setcounter{equation}{0}

In relativistic quantum field theory, the states of the quantum fields are defined on Cauchy surfaces, which one usually takes to be spacelike hyperplanes.  A Lorentz observer is characterized by the family of hyperplanes containing events simultaneous to the observer at the ``same'' time, which are parallel to the simultaneity hyperplane at a fixed given time, $x^0 = 0$, say.  The dynamical evolution determines how the physical state changes as one moves from one spacelike hyperplane to the next.  This is the ``instant form'' of the dynamics in the language of \cite{Dirac:1949cp}.
The Poincar\'e transformations preserving the foliation slice by slice are the kinematical transformations (spatial translations and spatial rotations), while the other Poincar\'e generators are dynamical (and called the ``Hamiltonians'' in \cite{Dirac:1949cp}). 

In the presence of gravitation, foliations by spacelike hyperplanes are not available.  In the asymptotically flat case, however, this structure appears at infinity \cite{Fock}, a fact made particularly clear in the Hamiltonian formulation of general relativity \cite{Dirac:1958sc,Dirac:1958jc,Arnowitt:1962hi}.  The Poincar\'e group structure in Dirac's ``instant form'' was exhibited in the pioneering paper \cite{Regge:1974zd}, where precise boundary conditions at spatial infinity were given and shown to yield the Poincar\'e algebra as asymptotic symmetry algebra.

The asymptotic structure at null infinity was studied in
\cite{Bondi:1962px,Sachs:1962wk,Sachs:1962zza,Newman:1962cia,Penrose:1962ij,Penrose:1965am}
and shown to be invariant under an infinite dimensional algebra now called the
``BMS'' algebra (for a recent  review, see
\cite{Madler:2016xju,Alessio:2017lps}).   The enlargement of this algebra by
``super-rotations'' was more recently performed in
\cite{Banks:2003vp,Barnich:2009se,Barnich:2010eb,Barnich:2011ct,Barnich:2011mi,Barnich:2013axa,Barnich:2016lyg,Flanagan:2015pxa,Strominger:2016wns,Barnich:2017ubf}
where the Lorentz algebra is extended to the conformal algebra in 2
dimensions while even bigger enlargements where also proposed in
\cite{Campiglia:2014yka,Campiglia:2015yka,Campiglia:2016jdj,Campiglia:2016efb}.
The remarkable potential physical implications of the BMS algebra both for the
infrared structure of gravity
\cite{Strominger:2013jfa,He:2014laa,Cachazo:2014fwa,Strominger:2014pwa,Pasterski:2015tva,Campiglia:2015kxa,Compere:2016jwb,Conde:2016rom} and
for black hole physics
\cite{Hawking:2016msc,Compere:2016hzt,Compere:2016gwf,Hawking:2016sgy,Bousso:2017dny,Strominger:2017aeh,Bousso:2017rsx} have attracted considerable interest in the last years \cite{Strominger:2017zoo}.  

It is implicit in this exciting work that the BMS algebra is realized in the quantum theory in terms of charges acting in the Hilbert space of states of the theory. 
These charges should have an expression at spatial infinity in the ADM formulation of the evolution based  on foliations that become asymptotically parallel hyperplanes, corresponding to inertial observers at infinity.  However, the  boundary conditions adopted in \cite{Regge:1974zd} at spatial infinity to make the angular momentum finite also make all BMS charges identically vanishing. Technically, as shown in \cite{Regge:1974zd}, this is a consequence of the so-called parity conditions imposed on the leading order of the metric and its conjugate momentum as one recedes to spatial infinity.  In order to resolve 
this tension between the asymptotic structure at spatial infinity and the BMS algebra emerging at null infinity, one must adopt boundary conditions at spatial infinity different from those of \cite{Regge:1974zd}.

One cannot just drop the standard parity conditions, since the symplectic structure, the angular momentum and the ``boost charges'' generically diverge logarithmically without them \cite{Beig:1987zz}.  One must therefore find alternative conditions that preserve finiteness and, at the same time, leave room for a well-defined and non trivial action of the BMS algebra.  

We propose in this paper new boundary conditions at spatial infinity that fulfill this purpose.  These boundary conditions  (i) are invariant under the BMS algebra,  (ii)  make the symplectic form finite, (iii)  contain the Schwarzchild solution, the Kerr solution and their Poincar\'e transforms, and (iv)  make the Hamiltonian generators of the asymptotic symmetries integrable,  well-defined (finite) and generically non-zero.  

The new consistent boundary conditions given here involve parity conditions of a different type than those of \cite{Regge:1974zd}. The existence of alternative parity conditions making the symplectic structure finite was observed in the insightful work \cite{Compere:2011ve}, but their full consistency was not studied.  The work \cite{Compere:2011ve} went indeed in a somewhat orthogonal direction since it was concerned with relaxing the parity conditions altogether and dealing with the ensuing divergences through holographic renormalisation.  Nevertherless, the analysis of \cite{Compere:2011ve} and the subsequent developments of  \cite{Troessaert:2017jcm} on the structure of its asymptotic symmetry, were important for arriving at the new boundary conditions proposed in this paper.  

Our work is organized as follows.  In Section \ref{sec:hamasymcond}, we recall some classic background information on the ADM Hamiltonian treatment of asymptotically flat spacetimes with the parity conditions of \cite{Regge:1974zd}.  This is necessary to motivate and derive our results. We formulate the asymptotic conditions both in asymptotically cartesian and asymptotically spherical coordinates, as it turns out that the new boundary conditions are most conveniently expressed in asymptotically spherical coordinates. Next, in Section \ref{sec:NewBoundaryConditions}, we give the explicit form of the new boundary conditions and verify that they consistently contain the Schwarzschild solution, the Kerr solution, and their Poincar\'e transforms. We also work out the form of the asymptotic symmetries.  In Section \ref{sec:finiteness}, we prove that the Hamiltonian generators of the asymptotic symmetries are integrable and finite.  We also point out that the charges associated with supertranslations need not vanish. Section \ref{sec:BMSAlgebra} is devoted to showing that the Hamiltonian generators of the asymptotic symmetries close according to the BMS algebra. Finally, Section
\ref{sec:Conclusions} summarizes our results and comments on various possible directions for extending them.  Three technical appendices complete our paper.

We focus here on vacuum gravity. Furthermore, our analysis is carried out in the Hamiltonian formalism of \cite{Dirac:1958sc,Dirac:1958jc,Arnowitt:1962hi} throughout.  The boundary conditions are expressed on the canonical variables ``$(q,p)$'' at any given time.  The action to be used in the path integral is $\int (p\dot{q} - H) dt$ where $q(t)$ and $p(t)$ fulfill at all $t$'s the boundary conditions given in this paper  but are not assumed to obey the equations of motion.  There are also Lagrange multipliers in the action, which must define asymptotic symmetries, i.e., define transformations that preserve the boundary conditions.

\section{Background}
\label{sec:hamasymcond}

\subsection{Fall-off at spatial infinity -- RT parity conditions}
Our starting point are the standard Hamiltonian boundary conditions for asymptotically flat spacetimes, given on spatial slices that asymptote hyperplanes equipped  with asymptotically cartesian coordinates $x^i = (x,y,z)$ at spatial infinity ($r \rightarrow \infty$ with $r^2 = x^i x_i$).  On any such hypersurface, the spatial metric $g_{ij}$ and its conjugate momentum $\pi^{ij}$ behave as
\begin{flalign}
	\label{eq:RTasympI}
	g_{ij} &= \delta_{ij} + \frac{1}{r} \xbar h_{ij} + \frac{1}{r^2}
	h^{(2)}_{ij} + o(r^{-2}),\\
	\label{eq:RTasympII}
	\pi^{ij} &= \frac{1}{r^2} \xbar\pi^{ij} + \frac{1}{r^3} \pi^{(2)ij} +
	o(r^{-3}) \, .
\end{flalign}
Indices are lowered and raised with the background flat metric $\delta_{ij}$ and its inverse.  The coefficients in both expansions are functions on the unit sphere.  We adopt the general convention  that barred quantities, such as $\xbar h_{ij}$ or $\xbar\pi^{ij} $ are functions on the unit sphere and so are $O(1)$. The boundary conditions include the Schwarzschild and Kerr metrics.

Under a deformation of the constant time hypersurface parametrized by $(\xi^\perp \equiv \xi, \xi^i)$,  the canonical variables transform as \cite{Dirac:1958sc,Arnowitt:1962hi}
\begin{eqnarray}
\delta g_{ij} &=& 2 \xi g^{-\frac12} \left(\pi_{ij} - \frac12 g_{ij} \pi \right) +\cL_{\xi} g_{ij} \\
\delta \pi^{ij} &=& - \xi g^{\frac12}\left(R^{ij} - \frac12 g^{ij} R \right) + \frac12 \xi g^{-\frac12} \left(\pi_{mn} \pi^{mn} - \frac12 \pi^2 \right)  \nonumber \\
&& -2 \xi g^{-\frac12} \left(\pi^{im} {\pi_{m}}^j - \frac12 \pi^{ij} \pi \right) + g^{\frac12} \left(\xi^{\vert ij} - g^{ij} {\xi^{\vert m}}_{\vert m} \right) \nonumber \\
&& + \cL_{\xi} \pi^{ij}
\end{eqnarray}
where $\cL_\xi g_{ij}$ and $\cL_\xi \pi^{ij}$ are respectively the Lie derivatives of $g_{ij}$ and $\pi^{ij}$ along the vector field $\xi^{i}$,
\begin{eqnarray}
\cL_\xi g_{ij} &=& \xi_{i \vert j} + \xi_{j \vert i} \\
\cL_\xi g_{ij} &=& \left(\pi^{ij} \xi^m \right)_{\vert m} - {\xi^i}_{\vert m} \pi^{mj} - {\xi^j}_{\vert m} \pi^{im}
\end{eqnarray}

These boundary conditions are invariant under hypersurface deformations $(\xi^\perp \equiv \xi, \xi^i)$ that behave asymptotically as \cite{Regge:1974zd}
\begin{eqnarray}
 \xi &=& b_i x^i + a(\mathbf{n}) + O\left(r^{-1}\right) \label{eq:As1}\\
 \xi^i &=&{ b^i}_j x^j + a^i(\mathbf{n})+ O\left(r^{-1}\right) \label{eq:As2}
\end{eqnarray}
where $b_i$ and $b_{ij} = -b_{ji}$ are arbitrary constants while
$a(\mathbf{n})$ and $a^i(\mathbf{n})$ are arbitrary functions on the unit
sphere ($\mathbf{n}^i = \frac{x^i}{r}$). The constants $b_i$ parametrize the Lorentz
boosts (the corresponding term $- b^i x^0$ in $\xi^i$ can be absorbed in $a^i$
at any given time), whereas the antisymmetric constants $b_{ij} = -b_{ji}$
parametrize the spatial rotations.  The zero modes $a_0$ and $a_0^i$ of $a$
and $a^i$ are standard translations.  General functions $a$ and $a^i$ describe
``angle-dependent'' translations.  The boundary conditions (\ref{eq:RTasympI})
and (\ref{eq:RTasympII}) are therefore invariant under an asymptotic algebra
that has the Poincar\'e algebra as a subalgebra.  We note that with
(\ref{eq:RTasympI}) and (\ref{eq:RTasympII}), the constraints have the
following fall-off,
\be
\mathcal H  = O(r^{-3}), \qquad  \mathcal H_i = O(r^{-3}) \label{eq:ConstFallOff}
\ee
(in asymptotically Cartesian coordinates). 

In addition to containing the Schwarzschild and Kerr solutions and being invariant under (at least) the Poincar\'e transformations, consistent boundary conditions should fulfill two addition requirements:
\begin{itemize}
\item The surface integrals yielding the charges associated with the asymptotic symmetries should be finite and ``integrable''.  By ``integrable'', one means that the variation of the surface charge, which is a one-form in field space obtained from the bulk generator through integration by parts \cite{Regge:1974zd}, is exact.
\item The kinetic term ``$p \dot{q}$'', i.e., $\int d^3 x \, \pi^{ij} \dot{g}_{ij}$, should be finite, i.e., the symplectic structure should be well-defined.
\end{itemize}
The general boundary conditions (\ref{eq:RTasympI}) and (\ref{eq:RTasympII}) fail on both accounts.  For that reason, they must be strengthened, but in way that does not eliminate the Schwarzschild or Kerr solutions and keeps the Poincar\'e transformations among the asymptotic symmetries.

The parity conditions given  in \cite{Regge:1974zd} fulfill all the consistency requirements.  These parity conditions are extra conditions on the leading terms in the expansion  (\ref{eq:RTasympI}) and (\ref{eq:RTasympII}), which are requested to fulfill definite parity properties under the antipodal map $x^k \rightarrow -x^k$.  Explicitly:
\begin{equation}
	\label{eq:parityconditons}
	\xbar h_{ij}(-n^k) = \xbar h_{ij}(n^k), \qquad \xbar \pi^{ij}(-n^k) =
	-\xbar \pi^{ij}(n^k) \, .
\end{equation} These 
parity conditions are obeyed by the Schwarzchild and Kerr solutions.  They are invariant under the transformations (\ref{eq:As1}) and (\ref{eq:As2}) provided $a-a_0$ and $a^i-a^i_0$ are odd functions of $\mathbf{n}^i$, and thus are in particular invariant under the Poincar\'e algebra.  They play  a central role in the mathematical work \cite{Corvino:2003sp,Huang:2010yd,Corvino:2011js}.

The parity conditions of \cite{Regge:1974zd} make the kinetic term finite since the coefficient of the leading logarithmic singularity in 
\be
\int d^3 x  \,  \pi^{ij} \dot{ h}_{ij} =  \int \frac{dr}{r} \int \sin \theta d \theta d \varphi \,  \xbar \pi^{ij} \dot{\xbar h}_{ij} + \cdots  \label{eq:KTFinite}
\ee
 actually vanishes. Indeed,   the term $\xbar \pi^{ij} \dot{\xbar h}_{ij}$ is an odd function on the sphere, so that its integral over the sphere is zero.  The remaining terms in (\ref{eq:KTFinite}), denoted by dots, are finite since their integrands decrease strictly faster than $r^{-1}$. The parity conditions also render the Poincar\'e charges finite and integrable \cite{Regge:1974zd}.  However, the charges associated with the remaining angle-dependent translations are then found to be identically zero (except the spacetime momentum associated with the zero modes), so that the actual asymptotic symmetry algebra, obtained by taking the quotient of all the asymptotic symmetries by the pure gauge ones -- i.e., the ones with zero charges \cite{Benguria:1976in} --, is the finite-dimensional Poincar\'e algebra.  There is no room for the full BMS algebra with the parity conditions of  \cite{Regge:1974zd}.

\subsection{Spherical coordinates}

\subsubsection*{Boundary conditions}

It turns out that an alternative strenghtening of the boundary conditions exists, which is also consistent, but which admits the full BMS algebra as asymptotic symmetry algebra.  These boundary conditions  are based on different parity conditions and do not eliminate solutions with non-vanishing BMS charges.

To describe this alternative strenghtening of the boundary conditions, it is convenient to use spherical coordinates $(r, x^A)$ where $x^A$ are coordinates on the sphere. In these coordinates, the asymptotic conditions (\ref{eq:RTasympI}) and (\ref{eq:RTasympII}) read
\begin{flalign}
	g_{rr} &= 1 + \frac{1}{r} \xbar h_{rr} + \frac{1}{r^2} h^{(2)}_{rr} +
	o(r^{-2}), \label{eq:SCasymp1}\\
	g_{rA} &=  \frac{1}{r} h^{(2)}_{rA} +
	o(r^{-1}), \label{eq:grA}\\
	g_{AB} &= r^2 \xbar\gamma_{AB} + r \xbar h_{AB} +  h^{(2)}_{AB} +
	o(1), \label{eq:SCasymp3}\\
	\pi^{rr} &= \xbar \pi^{rr} + \frac{1}{r} \pi^{(2)rr} +
	o(r^{-1}), \label{eq:SCasymp4} \\
	\pi^{rA} &= \frac{1}{r} \xbar \pi^{rA} + \frac{1}{r^2} \pi^{(2)rA} +
	o(r^{-2}), \label{eq:SCasymp5}\\
	\pi^{AB} &= \frac{1}{r^2} \xbar \pi^{AB} + \frac{1}{r^3} \pi^{(2)AB}+
	o(r^{-3}), \label{eq:SCasymp6}
\end{flalign}
where $\xbar \gamma_{AB}$ is the unit metric on the sphere.  There can in fact
be $O(1)$-terms $\xbar h_{rA}$ in the metric coefficients $g_{rA}$ in (\ref{eq:grA}),  but we have assumed them to vanish. 
The leading terms in $g_{rA}$  can indeed always be set to zero by a change of coordinates of the form
\begin{equation}
	r' = r + o(r^0), \quad {x'}^A = x^A + \frac{1}{r} \tilde X^A (x^B) 
	+ o(r^{-1}).
\end{equation}
The Schwarzchild and Kerr solutions fulfill the condition \eqref{eq:grA},
which is preserved under Poincar\'e transformations (see below).  It is only
under this condition that we shall develop  the formalism.  Difficulties with
integrability of the charges arise when the $O(1)$-terms $\xbar h_{rA}$ in the metric coefficient $g_{rA}$ do not vanish, but we have not investigated them here since these terms do not appear to carry physical information, at least for the known solutions.  A similar stronger-than-expected fall-off of the mixed radial-angular components of the metric was imposed for asymptotically anti-de Sitter spacetimes in \cite{Henneaux:1985tv}, or in the hyperbolic description of \cite{Compere:2011ve}.

It is convenient for later purposes to trade the variable $g_{rr}$ for $\lambda \equiv \frac{1}{\sqrt{g^{rr}}}$, the asymptotic expansion of which is 
\be
\lambda = 1 + \frac{1}{r} \xbar \lambda +\frac{1}{r^2} \lambda^{(2)}
	+  o(r^{-2}),
\ee
with
\be
	\xbar \lambda = \frac{1}{2} \xbar h_{rr}. 
\ee
Similarly, we introduce
\be
\xbar k^A_B = \frac{1}{2} \xbar h^A_B + \xbar \lambda \delta^A_B\, , \qquad \xbar k = \xbar k^{AB} \xbar \gamma_{AB} \, .
\ee
The functions $\xbar k_{AB}$ on the sphere have the following geometrical meaning. Let $K_{AB}$ be the extrinsic curvature of the 2-spheres $r =$ constant. If one expands $K^A_{B}$ asymptotically, one gets (see Appendix \ref{sec:radialsplit})
\be
	K^A_B = - \frac{1}{r} \delta^A_B + \frac{1}{r^2} \xbar k^A_B +
	\frac{1}{r^3} {k^{(2)}}^A_B + o(r^{-3}),
\ee
i.e., $ \xbar k^A_B$ is the coefficient of the leading perturbation to $K^A_B$ from its background value $- \frac{1}{r} \delta^A_B$.

\subsubsection*{Asymptotic symmetries}

In polar coordinates, the  transformations that preserve the above boundary conditions have the following  behaviour at infinity:
\begin{gather}
	\label{eq:lorentzparamI}
	\xi = r b + f + O(r^{-1}), \quad \xi^r = W + O(r^{-1}),\quad \xi^A =
	Y^A + \frac{1}{r} I^A + O(r^{-2}),\\
	\label{eq:lorentzparamII}
	\xbar D_A \xbar D_B b + \xbar \gamma_{AB} b = 0, \quad \cL_Y \xbar
	\gamma_{AB} = 0,
\end{gather}
where $I^A$ is given in terms of $b$ and $W$ as
\begin{equation} \label{eq:EqForI}
	 I^A = \frac{2b}{\sqrt {\xbar\gamma}}
	\xbar\pi^{rA}  + \xbar D^A W.
\end{equation}
Here, $b, f, W, Y^A$ and $I^A$ are functions and vector fields defined on the
sphere, and $\xbar D_A$ is the covariant derivative associated with the unit metric $\xbar \gamma_{AB}$ on the sphere.

A few comments are in order:
\begin{itemize}
\item The function $b$ describes the Lorentz boosts.  Explicitly, in terms of the cartesian parameters $b_i$, one has
\be
b = b_1 \sin \theta \cos \varphi + b_2 \sin \theta \sin \varphi + b_3 \cos \theta,
\ee
which is the general solution of $\xbar D_A \xbar D_B b + \xbar \gamma_{AB} b = 0$.
\item The vectors $Y^A$ describe the spatial rotations and are the standard Killing vectors on the sphere,
\be
\hspace{-.5cm} Y = m^1 \left(-\sin \varphi \frac{\partial}{\partial \theta} - \frac{ \cos \theta}{\sin \theta} \cos \varphi \frac{\partial}{\partial \varphi}\right)+ m^2 \left(\cos \varphi \frac{\partial}{\partial \theta} - \frac{ \cos \theta}{\sin \theta} \sin \varphi \frac{\partial}{\partial \varphi}\right) + m^{3} \frac{\partial}{\partial \varphi}
\ee
\item $f$ contains the time translation through its zero mode $f_0$ ($f \equiv a$ in the above parametrization); the other modes define transformations  outside the Poincar\'e algebra.
\item $W$ contains the spatial translations.  In an expansion in terms of spherical harmonics ${Y^\ell}_m$, the translations are the spin-1 part, $W_P=\sum_{m=-1}^1 P^m {Y^1}_{m}(x^A)$.  One has
\begin{eqnarray}
\frac{\partial}{\partial x} &=&\sin \theta \cos \varphi \frac{\partial}{\partial r} + \frac1r \cos \theta \cos\varphi \frac{\partial}{\partial \theta} - \frac1r \frac{\sin \varphi}{\sin \theta} \frac{\partial}{\partial \varphi}, \\
\frac{\partial}{\partial y} &=& \sin \theta \sin \varphi \frac{\partial}{\partial r} + \frac1r \cos \theta \sin \varphi \frac{\partial}{\partial \theta} + \frac1r \frac{\cos \varphi}{\sin \theta} \frac{\partial}{\partial \varphi}, \\
\frac{\partial}{\partial z} &=& \cos \theta \frac{\partial}{\partial r} - \frac1r \sin \theta  \frac{\partial}{\partial \theta}
\end{eqnarray}
and one easily sees that the corresponding vectors $I^A_P$ on the unit sphere
fulfill $I^A _P= \xbar D^A W_P$.  Furthermore, $\xbar D^A I^B_P + \xbar D^B
I_P^A 
	+2 W_P \xbar\gamma^{AB} = 0$.
\item The equation $I^A = \frac{2b}{\sqrt {\xbar\gamma}} \xbar\pi^{rA}  + \xbar D^A W$
follows from the preservation of the condition $\xbar h_{rA} = 0$ on the leading order of $ g_{rA}$.
As we have seen, it is  fulfilled by the spatial translation Killing vectors of the flat metric, which has indeed $ g_{rA}= 0$ (to all orders).
\end{itemize}

\subsubsection*{Standard parity conditions}

In polar coordinates, the parity conditions of \cite{Regge:1974zd} read, in terms of coordinates on the unit sphere for which the antipodal map is $x^A \rightarrow - x^A$,
\begin{equation}
			\xbar h_{rr} \sim \xbar \pi^{rA} \sim \xbar h_{AB}= \text{even, } \qquad
			\xbar \pi^{rr} \sim \xbar \pi^{AB} =
			\text{odd.} \qquad \text{(R-T)}
\end{equation}
This implies
\be  \xbar \lambda \sim \xbar k_{AB} = \text{even.} \qquad \text{(R-T)}
\ee
In terms of the traditional coordinates ($\theta$, $\varphi$) for which the antipodal map is $\theta \rightarrow \pi - \theta$, $\varphi \rightarrow \varphi + \pi$, this is equivalent to
\begin{eqnarray}
&&	\xbar h_{rr} \sim \xbar \pi^{r\theta} \sim \xbar \pi^{\theta \varphi}\sim \xbar h_{\theta \theta}\sim \xbar h_{\varphi \varphi} = \text{even, } \qquad \text{(R-T)} \\
&&			\xbar \pi^{rr} \sim  \xbar \pi^{r\varphi} \sim  \xbar \pi^{\theta \theta}  \sim \xbar \pi^{\varphi\varphi} \sim \xbar h_{\theta \varphi} =
			\text{odd.} \qquad \text{(R-T)}
\end{eqnarray}

The leading divergence in the kinetic term reads
\be
  \int \frac{dr}{r} \int  d \theta d \varphi \,   \left( \xbar \pi^{rr} \dot{\xbar h}_{rr}  + \xbar \pi^{AB} \dot{\xbar h}_{AB} \right) \label{eq:KTRTFinite}
\ee
and vanishes with the R-T parity conditions.  The surfaces charges are also finite \cite{Regge:1974zd}.  The transformations that preserve the R-T boundary conditions are 
\be
f-f_0 = \text{odd}, \qquad W-W_P = \text{even.} \qquad \text{(R-T)}
\ee
It is because the arbitrary functions occuring in $f$ and $W$ ($f-f_0$ and $W- W_P$, respectively) have parity opposite to that of the translations that they have identically vanishing surface charges and that there is no room for the BMS symmetry with the parity conditions of \cite{Regge:1974zd}.

\section{New boundary conditions}
\label{sec:NewBoundaryConditions}

\subsection{Explicit form}

As we have annouced, there is a different way to achieve finiteness of both the kinetic term and of the surface charges, without making the BMS charges identically zero.  This alternative way involves as a key ingredient the imposition of different parity conditions, preserved under surface deformations for which $f$ and $f_0$ have same (even) parity, as well as $W$ and $W_P$, which are both odd.  These alternative boundary conditions are of mixed type, in the sense that spherical and radial projections of the metric have different parities.

To formulate the new conditions in a simple way, we make the change of
variables adapted to the description of the extrinsic geometry of the spheres
$r =$ constant.  That is, we make the change of variables $\xbar h_{rr}, \xbar
h_{AB} \rightarrow \xbar \lambda, \xbar k_{AB}$, extended to the conjugate momenta so as to preserve the kinetic term,
\be
 \int  d \theta d \varphi \,   \left( \xbar \pi^{rr} \dot{\xbar h}_{rr}  + \xbar \pi^{AB} \dot{\xbar h}_{AB} \right)
 =  \int  d \theta d \varphi \,   \left( \xbar p \dot{\xbar \lambda}  + \pi^{AB}_{(k)} \dot{\xbar k}_{AB} \right). \label{eq:KineticFinite}
 \ee
One finds
\begin{eqnarray}
&& \xbar \lambda = \frac12 h_{rr}, \qquad  \xbar k_{AB} =  \frac12 \xbar h_{AB} + \xbar \lambda \xbar \gamma_{AB}, \qquad \\
&& \xbar p = 2 \left(\xbar \pi^{rr} - \xbar \pi^A_A \right) , \qquad \pi^{AB}_{(k)} = 2 \xbar \pi^{AB} \, .
\end{eqnarray}

The set of parity conditions on the boundary values proposed in this paper are
\begin{equation}
			\xbar \lambda \sim \xbar \pi^{AB}= \text{even}, \qquad
			\xbar p \sim \xbar k_{AB} \sim \xbar \pi^{rA} =
			\text{odd},  \label{eq:NewPCond}
\end{equation}
or in terms of $(\theta,\varphi)$-components
\begin{eqnarray}
&&	\xbar \lambda \sim \xbar \pi^{r\varphi} \sim  \xbar \pi^{\theta \theta}  \sim \xbar \pi^{\varphi\varphi} \sim \xbar k_{\theta \varphi}
 = \text{even, }  \\
&&			\xbar p \sim   \xbar \pi^{r\theta} \sim \xbar \pi^{\theta \varphi}\sim \xbar k_{\theta \theta}\sim \xbar k_{\varphi \varphi} =
			\text{odd.} 
\end{eqnarray}
Because the variables  $(\xbar \lambda, \xbar p)$ and $(\xbar k_{AB}, 2 \xbar \pi^{AB})$ in each conjugate pair have opposite parities, the coefficient (\ref{eq:KineticFinite}) of the divergent piece in the kinetic term vanishes.  The parity of $\xbar \pi^{r A}$ does not matter in this argument since $\xbar h_{rA} = 0$.
 
The Schwarschild solution obeys the new parity conditions provided one redefines the radial coordinate $r$, $r \rightarrow r' = r(1 - \frac{m}{r})$, which has the effect of making $\xbar k_{AB} =0$ (and thus odd). The Kerr solution also obeys these parity conditions after the same radial coordinate transformation is made, because the term $\pi^{r \varphi}$  related to the rotations is subleading:  its $O(r^{-1})$ piece $\xbar \pi^{r \varphi}$ vanishes and obeys thus trivially both the R-T parity condition (even) or the new ones (odd). The Taub-NUT solution \cite{Misner:1963fr}, however, has a non-vanishing $\xbar \pi^{r \varphi}$  which may be taken to be even and to obey therefore the R-T parity conditions \cite{Bunster:2006rt}.   It is excluded by the new boundary conditions.  This does not mean that one cannot handle the Taub-NUT solution, but rather that it corresponds to a different sector that has to be treated separately.  With the new boundary conditions, the Taub-NUT solution cannot be regarded as a standard asymptotically flat solution as in \cite{Bunster:2006rt} -- something in any case in line with the fact that it has a different topology.

\subsection{Constraints}

The new parity conditions  do insure finiteness of the symplectic form but do not insure by themselves cancelation of the divergent pieces in the boost charges and angular momentum, contrary to the parity conditions of \cite{Regge:1974zd}.  Therefore, they must be supplemented by further asymptotic restrictions in order to achieve finiteness of the charges.  These extra conditions are extremely mild.

With the boundary conditions (\ref{eq:SCasymp1})-(\ref{eq:SCasymp6}), the constraints have the fall-off (\ref{eq:ConstFallOff}), or, in spherical coordinates, $\mathcal H  = O(r^{-1})$, $\mathcal H_r = O(r^{-1})$, $\mathcal H_A = O(1)$.
The strengthening of the boundary conditions is simply that the leading divergences in the constraints should be absent, i.e., one must impose
\be
\mathcal H  = o(r^{-1}), \qquad  \mathcal H_r = o(r^{-1}),  \qquad  \mathcal H_A = o(1) \label{eq:ConstFallOff2}
\ee
Because the constraints transform among themselves under surface deformations, these extra conditions are consistent. Furthermore, they are very mild as announced,  since they of course hold on-shell and hence do not remove any solution. 

To recapitulate, the complete set of new boundary conditions proposed in this paper is (\ref{eq:SCasymp1})-(\ref{eq:SCasymp6}) with (\ref{eq:NewPCond}) and (\ref{eq:ConstFallOff2}).

 \subsection{Preservation under surface deformations}
 
The new boundary conditions are invariant under the surface deformations (\ref{eq:lorentzparamI}),
(\ref{eq:lorentzparamII}) and  (\ref{eq:EqForI}) provided the functions $f$ and $W$ on the sphere fulfill the following conditions:

 \begin{itemize}
 \item The function $f$ has the form
 \be 
 f = - 3 b \xbar \lambda - \frac12 b \xbar h + T \equiv - b \xbar \lambda -  b \xbar k + T  \label{eq:Functionf}
 \ee
 where $T$ is an arbitrary even function on the sphere,
 \be
 T = \text{even}, \label{eq:TEven}
 \ee
 \item The function $W$ is  an arbitrary odd function on the sphere,
 \be
 W = \text{odd}. \label{eq:WOdd}
 \ee
 \end{itemize}
The transformations that preserve the boundary conditions contain therefore the Poincar\'e transformations.  There are in addition arbitrary angle-dependent translations, but now these have the same parity as the ordinary translations.  We shall show in Section \ref{sec:finiteness} below that the corresponding charges are all integrable and finite.

The term $ - b \xbar \lambda -  b \xbar k$ must be included in $f$ to cancel terms with incorrect parity in the variation of the canonical variables.  For instance,  $\delta_\xi \xbar \pi^{rA}$ reads
\be
\delta_\xi \xbar \pi^{rA}  = \cL_Y \xbar\pi^{rA} + \sqrt{\xbar
	\gamma} \left(\xbar D_B(b \xbar k^{BA}) - b \xbar D^A \xbar k - \xbar
	D^A (\xbar \lambda b) - \xbar D^A f \right)
\ee
The term $- \xbar D^A (\xbar \lambda b) $  is odd, rather than being even to conform with the parity of  $\xbar \pi^{rA}$, and must therefore be cancelled by  $- \xbar D^A f$.  Together with the requirement of integrability of the charges discussed below, this forces $f$ to be given by (\ref{eq:Functionf}).
 
Under a transformation generated by the gauge parameters
\eqref{eq:lorentzparamI} and \eqref{eq:lorentzparamII} with $f = T -
b\xbar k - b\xbar\lambda$ and $I_A$ given by (\ref{eq:EqForI}), the components of 
the metric have the following behaviour
\begin{flalign}
	\delta_\xi \xbar k_{AB} & = \cL_Y \xbar k_{AB} + \xbar D_A \xbar D_B W + W
\xbar \gamma_{AB}\nonumber \\ &\qquad + \frac{b}{\sqrt{\xbar\gamma}} (\xbar \pi_{AB} - \xbar
\gamma_{AB} \xbar \pi) + \frac{1}{\sqrt{\xbar\gamma}} \xbar D_A( b \xbar
\pi^{rC} \xbar\gamma_{CB}) + \frac{1}{\sqrt{\xbar\gamma}} \xbar D_B( b \xbar
\pi^{rC} \xbar\gamma_{CA}),\\
	\delta_\xi \xbar\lambda &= \frac{b}{4\sqrt 
	{\xbar\gamma}} 
	\xbar p +  Y^C
	\d_C \xbar \lambda,
\end{flalign}
where $\xbar \pi = \xbar \pi^{AB} \xbar \gamma_{AB}$.
For the momenta, one has
\begin{flalign}
	\delta_\xi\xbar p & = \cL_Y \xbar p +\sqrt{ \xbar \gamma} \left( 4 b\xbar
	D_C \xbar D^C \xbar \lambda + 4 \xbar D^C b \d_C \xbar \lambda
+ 12 b \xbar \lambda\right) \\
	\delta_\xi \xbar \pi^{rA} & = \cL_Y \xbar\pi^{rA} + \sqrt{\xbar
	\gamma} \left(\xbar D_B(b \xbar k^{BA}) + \xbar D^A b \xbar k
	- \xbar D^A T\right), \\
	\delta_\xi \xbar \pi^{AB} & = \cL_Y \xbar \pi^{AB} +
	\sqrt{\xbar\gamma} \left( \xbar D^A \xbar D^B T - \xbar \gamma^{AB}
		\xbar D_C \xbar D^C T\right) + 3b \sqrt{\xbar\gamma}\left(\xbar k^{AB} - \xbar
	\gamma^{AB} \xbar k\right) \nonumber \\ & \qquad + \sqrt{\xbar\gamma} b
	\left(\xbar\gamma^{AB} \xbar D_C \xbar D^C \xbar k + \xbar D_C
		\xbar D^C \xbar k^{AB} - \xbar D_C \xbar D^A \xbar
	k^{CB} - \xbar D_C \xbar D^B \xbar k^{CA}\right)\nonumber\\ &
	\qquad +\sqrt{\xbar \gamma} \Big( - \xbar D^A b \xbar D^B
	\xbar k - \xbar D^B b \xbar D^A \xbar k+ \xbar \gamma^{AB} \xbar D_C
	b \xbar D^C \xbar k  + 2 \xbar \gamma^{AB} \xbar D^D \xbar k^C_D \d_C
	b\nonumber\\ & \qquad \qquad - \xbar D^A \xbar k^{BC} \d_Cb-
\xbar D^B \xbar k^{AC} \d_Cb + \xbar D^C \xbar k^{AB} \d_C b\Big).
\end{flalign}
In order to obtain the transformation law of $\xbar p$, we used the identity 
\begin{equation}
	\xbar D^A \xbar D^B \xbar k_{AB} - \xbar D_A \xbar D^A \xbar k = 0,
\end{equation}
coming from the extra condition $\mathcal H=o(r^{-1})$ (see appendices~\ref{sec:radialsplit} and~\ref{app:loga}).

One can verify from these formulas that the parity conditions are all preserved by the surface deformations.  Note that for the boosted Schwarzschild metric in the coordinates for which $\xbar k_{AB} = 0$, the only component of the momentum that acquires a non vanishing value is $\xbar p$, which is correctly odd and equal in this case to $\xbar \pi^{rr}$.

\section{Asymptotic charges}
\label{sec:finiteness}

We now show that the canonical generators of the asymptotic symmetries are well-defined with the new parity conditions. We follow the method of \cite{Regge:1974zd} and do not impose $\xbar h_{rA} =0$ to begin with.  This will be done later, at the point where it is needed. We set $\lambda_A \equiv g_{rA}$.

Our aim is to show that the bulk piece of the generators, given by the smeared constraints $ \int d^3x \, \left(\xi \mathcal H + \xi^i \mathcal H_i \right)$, can be supplemented by appropriate surface terms that make the sum ``differentiable'' when  $\xi$ and $\xi^i$ are given by (\ref{eq:lorentzparamI}), (\ref{eq:lorentzparamII}), (\ref{eq:EqForI}) and (\ref{eq:Functionf}), where $T$ and $W$ are arbitrary field-independent even and odd functions, respectively.  The boost and rotation parameters $b$ and $Y^A$ are also taken to be field independent.

Taking a general variation of the smeared constraints, we obtain:
\begin{equation}
	\delta \int d^3x \, \left(\xi \mathcal H + \xi^i \mathcal H_i
	\right) = \int d^3 x \left( 
	\delta_\xi \pi^{ij} \delta g_{ij}-\delta_\xi g_{ij} \delta
	\pi^{ij}\right) + \lim_{r\rightarrow \infty} \mathcal K_\xi[\delta
	g_{ij}, \delta \pi^{ij}],
\end{equation}
where the boundary term is given by
\begin{multline}
\mathcal K_\xi[\delta
	g_{ij}, \delta \pi^{ij}] = 
	\oint d^2x \, \Big \{ -2 \xi^i \delta \pi^r_i + \xi^r
		\pi^{ij} \delta g_{ij} - 2\sqrt \gamma \xi\delta K\\
	- \sqrt \gamma \gamma^{BC}\delta \gamma_{CA}\left(\xi K^A_B + \frac{1}{\lambda} (\d_r \xi -
	\lambda^D \d_D \xi)\delta^A_B\right) \Big\}.
\end{multline}
In order to write this term, we used a radial 2+1 split of the 3d metric
$g_{ij}$ (see appendix~\ref{sec:radialsplit} for more details, including conventions).  

Collecting all divergent and finite terms, we get
\begin{multline}
\mathcal K_\xi[\delta
	g_{ij}, \delta \pi^{ij}] = 
	r \oint d^2x \, \Big \{ -2 \, Y^A \xbar\gamma_{AB} \delta\xbar\pi^{rB}  -2\sqrt{\xbar \gamma} b \delta
		\xbar k\Big\} \\
			+\oint d^2x \Big\{-2 Y^A \delta (\xbar h_{AB} \xbar \pi^{rB} + \xbar \gamma_{AB}
	\pi^{(2)rB} + \xbar \lambda_A \xbar \pi^{rr}) -2 I^A \xbar\gamma_{AB} \delta \xbar
	\pi^{rB} \\- 2 W \delta \xbar \pi^{rr}- \sqrt{\xbar\gamma} (b \xbar h \delta \xbar k + 2 f \delta
	\xbar k + 2b \delta k^{(2)})\\
	+\sqrt{\xbar \gamma} ( f+ \xbar \lambda b + \xbar
	\lambda^D\d_D b )\delta \xbar h-\sqrt{\xbar \gamma} 
	b\, \xbar k^{AB}\delta \xbar h_{AB}  \Big\}+ o(r^0).
\end{multline}
If non-zero, the first term is divergent. This is what motivated the
introduction of the parity conditions (\ref{eq:parityconditons}) in \cite{Regge:1974zd}, which makes the potentially divergent term identically zero. 

Here, a different mechanism is at play. That is, the fact that the constraints hold asymptotically in the sense of (\ref{eq:ConstFallOff2}) is sufficient to remove the divergence (independently in fact of any parity condition).  In that sense, the parity condition of \cite{Regge:1974zd} ``kills twice'' the divergence.
Indeed, the leading
terms of the constraints take the form
\begin{gather}
	\mathcal H_A 
	= -2 \xbar \gamma_{AB} (\xbar \pi^{rB} +\xbar D_C \xbar
	\pi^{BC}) + o(1),\quad
	\mathcal H = -
	 \frac{2}{r} \sqrt{\xbar\gamma}\left(\xbar D_A \xbar D_B \xbar k^{AB} -
	\xbar D_A \xbar D^A \xbar k\right)  + o(r^{-1})
\end{gather}
(see Appendix \ref{app:loga} for more information). Since both these terms are equal to zero by (\ref{eq:ConstFallOff2}), we
can rewrite the divergent contribution as
\begin{multline}
\mathcal K_\xi[\delta
	g_{ij}, \delta \pi^{ij}] = 
	r\oint d^2x \, \Big \{ 2\, Y^A \xbar\gamma_{AB} \xbar
	D_C\delta\xbar\pi^{BC}\\  - 2\sqrt
	{\xbar\gamma} \, b \delta(\xbar k - \xbar D_A \xbar D_B \xbar k^{AB} + \xbar
	D_A \xbar D^A \xbar k )\Big\}+ O(1).
\end{multline}
If we then integrate by parts and use the properties of the
Lorentz parameters given in \eqref{eq:lorentzparamII}, we see that this divergence cancels.

Using the fact that both $Y^A$ and $b$ are field independent, we can partially
integrate the finite part of the boundary term:
\begin{multline}
\mathcal K_\xi[\delta
	g_{ij}, \delta \pi^{ij}] =
	\delta\oint d^2x \, \Big \{	-2 Y^A (\xbar h_{AB} \xbar \pi^{rB} + \xbar \gamma_{AB}
	\pi^{(2)rB} + \xbar \lambda_A \xbar \pi^{rr})
	\\- 2\sqrt{\xbar\gamma} b k^{(2)}  - \sqrt{\xbar\gamma} \frac{1}{4} b
	( \xbar h^2 + \xbar h^{AB} \xbar h_{AB})\Big\}\\
	+\oint d^2x \, \Big \{	 -2 I^A \xbar\gamma_{AB} \delta \xbar
	\pi^{rB} - 2 W \delta \xbar \pi^{rr}
	- \sqrt{\xbar\gamma}(2f +  \xbar h b) \delta (2\xbar \lambda
	+ \xbar D_A \xbar \lambda^A)\\ + \sqrt{\xbar\gamma} (\xbar \lambda^C\d_C
	b \,\xbar\gamma^{AB} - b\xbar D^A \xbar\lambda^B) \delta \xbar h_{AB}\Big\}+ o(r^0).
\end{multline}
To integrate the terms written in the last line, we now use the conditions $\xbar \lambda_A = 0$ and (\ref{eq:Functionf}).   Without restriction on these parameters, the one-form in field space  $\mathcal K_\xi$ is not exact\footnote{A different viewpoint on integrability has been recently developed in \cite{Troessaert:2015nia}.}.
We then get
\begin{equation}
	\mathcal K_\xi[\delta g_{ij}, \delta \pi^{ij}] = - \delta \mathcal
	B_\xi[g_{ij}, \pi^{ij}],
\end{equation}
where
\begin{multline}
	\label{eq:boundtermBgen}
	\mathcal B_\xi[g_{ij}, \pi^{ij}] = \oint d^2x \Big
	\{  
	Y^A \Big(4\xbar k_{AB} \xbar \pi^{rB} - 4 \xbar \lambda\xbar \gamma_{AB}
	\xbar\pi^{rB}+ 2 \xbar \gamma_{AB}
	\pi^{(2)rB}\Big) + W \Big( 2 \xbar \pi^{rr} - 2 \xbar D_A \xbar
	\pi^{rA}\Big) \\ + T\, 4\sqrt{\xbar\gamma}\,  \xbar
	\lambda +b\,
	\sqrt{\xbar \gamma} \Big( 2  k^{(2)} + \xbar k^2 +
	\xbar k^A_B \xbar k^B_A - 6 \xbar\lambda\xbar k\Big) +b\frac{2}{\sqrt{\xbar\gamma}} \xbar \gamma_{AB}
	\xbar\pi^{rA}\xbar\pi^{rB}\Big\}.
\end{multline}
Using the parity conditions \eqref{eq:NewPCond}, this boundary term can be simplified to
\begin{equation}
	\label{eq:boundtermB}
	\mathcal B_\xi[g_{ij}, \pi^{ij}] = \oint d^2x \Big
	\{T\, 4\sqrt{\xbar\gamma}\,  \xbar
	\lambda  + W\, \xbar p +
	Y^A\, 2 \xbar \gamma_{AB}\left( \pi^{(2)rB}- 2 \xbar \lambda
	\xbar\pi^{rB}\right)+b\,
	2 \sqrt{\xbar \gamma} \left( k^{(2)}  - 3 \xbar\lambda\xbar k\right)\Big\}.
\end{equation}
It is interesting to note that integrability alone of the term $\sqrt{\xbar\gamma}(2f +  \xbar h b) \delta (2\xbar \lambda)$ leaves some freedom in the choice of $f$, since any function of $\xbar \lambda$ could be added to $f$ without destroying integrability.  On the other hand, the preservation under surface deformations of the parity conditions controls, as we have seen,  the $\xbar \lambda$-dependence and removes the ambiguity.  It is very satisfying to see that both the integrability conditions and the parity conditions combine to fix the form of the charges. 

We can summarize the above results in the following theorem:
\begin{theorem}
The transformations associated with the asymptotic symmetries
\begin{gather}
	\xi =  b\Big(r -  \xbar \lambda -   \xbar k \Big) + T + O(r^{-1}),\quad \xi^A =
	Y^A + \frac{1}{r} \Big(\xbar D^A W + \frac{2b}{\sqrt{\xbar\gamma}}
\xbar \pi^{rA}\Big) + O(r^{-2}),\\
	\xi^r = W + O(r^{-1}), \quad \xbar D_A \xbar D_B b + \xbar \gamma_{AB} b = 0, \quad \cL_Y \xbar
	\gamma_{AB} = 0, \\
 T = \text{even function} , \qquad 
 W = \text{odd function},	
\end{gather}
where $b, Y^A, T$ and $W$ are field independent, are canonical
transformations generated by 
\begin{equation}
	G_{\xi}[g_{ij}, \pi^{ij}] = \int d^3x \, \left(\xi \mathcal H + \xi^i \mathcal H_i
	\right) + \mathcal B_\xi[g_{ij}, \pi^{ij}],
\end{equation}
where the boundary term $\mathcal B_\xi$ is given in 
equation~\eqref{eq:boundtermB}.
\end{theorem}
This shows that the asymptotic symmetries have well-defined (differentiable) generators.

\vspace{.2cm}
We close this section with four comments.
\begin{itemize}
\item As we pointed out, finiteness of the charges holds even without any parity condition and is a consequence of the asymptotic implementation of the constraints.  However, if one drops the parity conditions, there is some ambiguity in the $\xbar \lambda$-dependence of the boost generators (in addition to the singularity of the symplectic structure).
\item Time and spatial translations, 
respectively given by $T=1$ and $W_P=\sum_{m=-1}^1 P^m Y_{1m}(x^A)$, have
associated charges
\begin{equation}
	\mathcal M[g_{ij}, \pi^{ij}] = 4\oint d^2x 
	\sqrt{\xbar\gamma} \,  \xbar
	\lambda, \quad \mathcal P[g_{ij}, \pi^{ij}] = \sum_{m=-1}^1 P^m\, \oint d^2x 
	\,Y_{1m} \xbar p,
\end{equation}
where $\mathcal M$ is the mass while $\mathcal P$ the linear
momentum. These expressions agree
with the ADM ones, as they should. 
\item The situation is more subtle for the Lorentz
transformations parametrized by $b$ and $Y^A$. Their charges will in general
contain non-linear terms in the dynamical fields:
\begin{gather}
\label{eq:LorentzCharges1}
	\mathcal B_Y[g_{ij}, \pi^{ij}] = \oint d^2x \,
	Y^A\, 2 \xbar \gamma_{AB}\left( \pi^{(2)rB}- 2 \xbar \lambda
	\xbar\pi^{rB}\right),\\
	\mathcal B_b[g_{ij}, \pi^{ij}] = \oint d^2x\, b\,
	2 \sqrt{\xbar \gamma} \left( k^{(2)}  - 3 \xbar\lambda\xbar k\right). \label{eq:LorentzCharges2}
\end{gather}
When one uses the R-T parity conditions given in 
equation~\eqref{eq:parityconditons} in the general expression \eqref{eq:boundtermBgen}, all non-linear terms disappear and we recover 
the results of \cite{Regge:1974zd}.  But with the new parity conditions, some
of the non-linear terms are generically non-vanishing. This non-linearity is similar to what was encountered  for anti-de Sitter gravity coupled to a scalar field, either in the holographic renormalization approach \cite{Bianchi:2001de,Bianchi:2001kw}, or through canonical methods \cite{Henneaux:2002wm,Henneaux:2004zi,Henneaux:2006hk}.
\item The charges associated with supertranslations are 
\begin{equation}
	\mathcal B_\xi[g_{ij}, \pi^{ij}] = \oint d^2x \Big
	\{  T\, 4\sqrt{\xbar\gamma}\,  \xbar
	\lambda + W \xbar p \Big\}.
\end{equation}
and do not vanish in general since they are given by the integrals of non trivial even functions.
\end{itemize}

\section{The BMS algebra}
\label{sec:BMSAlgebra}

In order to compute the algebra of the asymptotic symmetries, we have to take into
account the fact that they have an explicit phase-space dependence. 
The resulting bracket
between two asymptotic transformations $\xi_1(Y_1, b_1, T_1, W_1)$ and 
$\xi_2(Y_2, b_2, T_2, W_2)$ is then given by
\begin{equation}
	[\xi_1, \xi_2]_M  = 
	[\xi_1,\xi_2]_{SD} + \delta_2^{g,\pi} \xi_1 - \delta_1^{g,\pi} \xi_2 +
	\Theta^A(\xi_1, \xi_2) \mathcal H_A + \Theta(\xi_1, \xi_2) \mathcal H,
\end{equation}
where $[,]_{SD}$ is the usual surface deformation bracket while the 
variations $\delta^{g,\pi}$ only hit the explicit dependence on the
gravitational fields. The extra terms proportional to the constraints contain
the contribution to the Poisson bracket produced when the Euler-Lagrange
derivatives only hit the gauge parameters. In this case, they can be ignored
safely. One way to see this is that the
improper part of the gauge parameters explicitly depends only on one element
from each canonical pair namely $g_{rr}$, $\pi^{rA}$ and $g_{AB}$. This
implies that non-zero $\Theta$ or $\Theta^A$ will always involve the proper part of
at least one of the two gauge parameters. As the bracket of a proper
gauge transformation with any allowed transformation is a proper gauge
transformation, these contributions are always sub-leading.

Using results for the variations of the asymptotic fields given in Section \ref{sec:NewBoundaryConditions}, the computation of the algebra of 
asymptotic transformations is straightforward:
\begin{equation}
	\hat\xi(\hat Y, \hat b,\hat T,\hat W) = \Big[\xi_1(Y_1, b_1, T_1,
		W_1), \xi_2(Y_2, b_2, T_2, W_2)\Big]_M,
\end{equation}
where
\begin{flalign}
	\label{eq:hamilbmsI}
	\hat Y^A & = Y^B_1\d_B Y_2^A + \xbar \gamma^{AB} b_1\d_B b_2 - (1
	\leftrightarrow 2),\\
	\label{eq:hamilbmsII}
	\hat b & = Y^B_1\d_B b_2 - (1 \leftrightarrow 2),\\
	\label{eq:hamilbmsIII}
	\hat T & = Y_1^A\d_A T_2 - 3 b_1 W_2 - \d_A b_1 \xbar D^A W_2 - b_1
	\xbar D_A\xbar D^A W_2 - (1 \leftrightarrow 2),\\
	\label{eq:hamilbmsIV}
	\hat W & = Y_1^A \d_A W_2 - b_1T_2 - (1 \leftrightarrow 2).
\end{flalign}
When the functions $T$ and $W$ are restricted to be respectively an even and
an odd function on the sphere, the bracket obtained here describes the BMS
algebra using an unfamiliar
basis. The proof can be found in Appendix~\ref{app:Hyperbolic}. The main idea is to relate the algebra
obtained here with the asymptotic analysis  performed in
\cite{Troessaert:2017jcm} in the context of the hyperbolic treatment of
spatial infinity developed in \cite{Ashtekar:1978zz,BeigSchmidt,Beig:1983sw}.
The result is a linear isomorphism between the set of pairs of functions 
$(T, W)$ with even and odd parity and the set of functions $\mathcal T$ of no definite parity
such that $\mathcal T$ transforms as a usual BMS supertranslation under the action 
of Lorentz algebra:
\begin{equation}
	\delta_{Y,b} \mathcal T = Y^A \d_A \mathcal T - \xbar D^A b \d_A
	\mathcal T - b \mathcal T.
\end{equation}
Developing the various functions in spherical harmonics, one can write
the first few terms of the change of basis:
\begin{gather}
	W = \sum_k \sum_{m=-2k-1}^{2k+1} W_{2k+1,m}\, Y_{2k+1,m}, \quad
	T = \sum_k \sum_{m = -2k}^{2k} T_{2k,m} \, Y_{2k,m},\\
	\mathcal T= T_{0,0} Y_{0,0} +
	\sum_{m=-1}^1 W_{1,m} Y_{1,m} +\frac 1 4 \sum_{m=-2}^2 T_{2,m} Y_{2,m} + ...
\end{gather}
where the dots denotes terms containing spherical harmonics with $l>2$.

General theorems guarantee that the generators associated to asymptotic
symmetries close in the Poisson bracket according to the same algebra, possibly modified by central charges \cite{Brown:1986ed}.   One can check that in the present case, however, the algebra  does not acquire a central extension:
\begin{equation}
	\Big\{G_{\xi_1}[g_{ij}, \pi^{ij}], G_{\xi_2}[g_{ij}, \pi^{ij}]\Big\} =
	G_{\hat\xi}[g_{ij}, \pi^{ij}].
\end{equation}
The easiest way to check this is to express the Poisson bracket as a variation
$\{G_{\xi_1}, G_{\xi_2}\} = \delta_{\xi_2} G_{\xi_1}$ and evaluate the result
on the background Minkowski space.

\section{Conclusions}
\label{sec:Conclusions}

In this paper, we have proposed new boundary conditions for asymptotically flat spacetimes at spatial infinity.  These new boundary are given by (\ref{eq:SCasymp1})-(\ref{eq:SCasymp6}), with the parity conditions (\ref{eq:NewPCond}) on the leading order of the asymptotic fields and the requested constraint fall-off  (\ref{eq:ConstFallOff2}).

These boundary conditions fulfill all the standard consistency requirements: they contain the Schwarzschild and Kerr solutions as well as their Poincar\'e transforms; the symplectic structure is well-defined; the generators of the asymptotic symmetries, which contain asymptotic Poincar\'e transformations, are all finite.  We have also constructed explicitly the conserved charges of the asymptotic symmetries and showed that they close according to the BMS algebra, which is consequently the asymptotic symmetry algebra.

We have therefore achieved the goal outlined in the introduction, of associating standard canonical  generators at spatial infinity to the BMS symmetry transformations first revealed at null infinity.     These generators do not identically vanish and hence have a non trivial action in the physical phase space.

A key ingredient in the new boundary conditions are the parity conditions (\ref{eq:NewPCond}), which are different from those proposed earlier in \cite{Regge:1974zd}.   We have seen that the BMS super-translations are encoded in the odd part of $W$ and the even part of $T$. Both of these parts are
incompatible with the parity conditions of  \cite{Regge:1974zd} and hence absent in that approach, except for the few spherical harmonics describing Poincar\'e translations. [The even part of $W$ and the odd part of $T$ are compatible with the boundary conditions of \cite{Regge:1974zd} but have zero charges because their parity is opposite to that of the Poincar\'e translations.  Hence they are pure gauge.]

By contrast, our new parity conditions allow arbitrary odd $W$'s and even $T$'s and a non trivial action of the BMS group at spatial infinity.  The new parity conditions are motivated by the behaviour of the gravitational field at null infinity and hence the behaviour of gravitational radiation.  The  parity conditions of \cite{Regge:1974zd} were instead motivated by the behaviour of boosted Schwarszchild in standard coordinates. It is quite remarkable that the  boosted Schwarszchild solution also fulfills the new parity conditions, provided one performs a rather straightforward change of coordinates.

Our work can be extended in various directions.  

\begin{itemize}
\item The superrotations \cite{Barnich:2009se,Barnich:2010eb,Barnich:2011ct,Barnich:2011mi,Barnich:2013axa,Barnich:2016lyg} are not included among the asymptotic symmetries described here. It would be of interest to examine if and  how they can be covered.
\item The new parity conditions do not include the Taub-NUT solution.  To cover it, one presumably needs to consider it as defining a distinct ``sector'' and study perturbations around it (``asymptotically Taub-NUT spacetimes'').  It would be useful to carry out the study explicitly.  This would need a more detailed analysis of the electric and magnetic components of the Weyl tensor at infinity.
\item Our paper focused on vacuum gravity. Matter fields, and most notably,
	the electromagnetic field, should be included. Concerning the latter,
	a first step has been done in \cite{Campiglia:2017mua} where the
	authors have realized at spatial infinity, using a hyperbolic slycing,
	a description of the
	enlarged asymptotic symmetry of eletromagnetism introduced in
	\cite{Strominger:2013lka,Barnich:2013sxa}. 
\item Finally, it would be of interest to investigate possible relaxations of the boundary conditions beyond parity conditions \cite{Compere:2011ve}.  We have seen that the asymptotic implementation of the constraints, without parity conditions, is sufficient to ensure finiteness of the charges but not of the canonical kinetic term $\int dt \, p \dot{q}$.    The authors of \cite{Compere:2011ve} use the framework of holographic renormalisation to remove the divergences that appear in the symplectic structure but end up with a puzzle that they lucidly describe in their conclusions: instead of a single phase space, they get a collection of phase spaces where some BMS transformations are not allowed to act. How this would translate in the ADM approach is worth pursuing.

In a related context, a different set of boundary conditions at spatial infinity having a BMS algebra as symmetry has been presented in \cite{Baghchesaraei2016}.  However, this representation of the BMS algebra does not contain spatial translations nor Lorentz boosts and, as such, is not the usual BMS algebra considered at null infinity.  Their analysis is nevertheless very interesting and it may hint at possible generalisations of the results presented in this work.  
\end{itemize}

It is hoped to return to these questions in the future.

\section*{Acknowledgments} 
This work was partially supported by the ERC Advanced Grant ``High-Spin-Grav", by FNRS-Belgium (convention FRFC PDR T.1025.14 and  convention IISN 4.4503.15) and by the ``Communaut\'e Fran\c{c}aise de Belgique" through the ARC program.

\begin{appendix}

\section{Radial decomposition of the spatial metric and the spatial curvature}

\label{sec:radialsplit}

Let us assume that we have spatial coordinates given by $x^i=(r,x^A)$
where $x^A$ are coordinates on the 2-sphere. We introduce:
\begin{equation}
\gamma_{AB} \equiv g_{AB}, \quad \lambda_A \equiv g_{rA}, \quad \lambda \equiv \frac{1}{\sqrt{g^{rr}}}.
\end{equation}
The metric and its inverse take the form:
\begin{equation}
g_{ij}= \left(\begin{array}{cc}
\lambda^2 + \lambda_C\lambda^C & \lambda_B \\
\lambda_A & \gamma_{AB}
  \end{array}
\right),\quad 
g^{ij}= \left(\begin{array}{cc}
\frac{1}{\lambda^2} & -\frac{\lambda^B}{\lambda^2} \\
-\frac{\lambda^A}{\lambda^2} & \gamma^{AB}+ \frac{\lambda^A\lambda^B}{\lambda^2}
  \end{array}
\right),
\end{equation}
where we used $\gamma_{AB}$ and its inverse $\gamma^{AB}$ to raise and
lower the angular indices $A, B, ...$

Introducing the extrinsic curvature of the 2-spheres $K_{AB}$, we can write
all the Christoffel symbols:
\begin{eqnarray}
K_{AB} & =  & \frac{1}{2 \lambda} \left( - \d_r g_{AB} + D_A \lambda_B
  + D_B \lambda_A\right)\\
\Gamma^r_{AB} & = & \frac{1}{\lambda} K_{AB} \\
\Gamma^A_{BC} & = & {}^\gamma\Gamma^A_{BC} - \frac{\lambda^A}{\lambda} K_{BC} \\
\Gamma^r_{rA} & = & \frac{1}{\lambda} \left( \d_A \lambda + K_{AB} \lambda^B\right) \\
\Gamma^r_{rr} & = &\frac{1}{\lambda} \d_r \lambda
+\frac{\lambda^A}{\lambda} \left( \d_A \lambda + K_{AB}
  \lambda^B\right) \\ 
\Gamma^A_{rB} & = & -\frac{\lambda^A}{\lambda} \left( \d_B \lambda + K_{BC}
  \lambda^C\right) + D_B \lambda^A - \lambda K^A_B \\
\Gamma^A_{rr} & = & -\lambda \left( \gamma^{AB} + \frac{\lambda^A
    \lambda^B}{\lambda^2}\right) \left( \d_B \lambda +
  K_{BC}\lambda^C\right) - \lambda^C \left( D^A \lambda_C - \lambda
  K^A_C\right) \nonumber \\ && \qquad- \frac{\lambda^A}{\lambda} \d_r \lambda + \gamma^{AB}
\d_r \lambda_B
\end{eqnarray}
where $D_A$ is the covariant derivative associated to $\gamma_{AB}$.

The Ricci tensor is given by:
\begin{eqnarray}
	{}^{(3)}R_{AB} & = & \frac{1}{\lambda} \d_r K_{AB} + 2K_{AC}K^C_B -
	K K_{AB} - \frac{1}{\lambda} D_AD_B \lambda \nonumber \\ 
	&& \quad + {}^{\gamma} R_{AB} - \frac{1}{\lambda} 
	\cL_\lambda K_{AB},\\
	{}^{(3)}R_{rA} & = & \lambda \left( \d_A K - D_B
	K^B_A\right) + {}^{(3)}R_{AB} \lambda^B,\\
	{}^{(3)}R_{rr} & = & \lambda(\d_rK - \lambda^A\d_A K) - \lambda^2
	K^A_B K^B_A - \lambda D_AD^A\lambda \nonumber \\ && \quad-
	{}^{(3)}R_{AB}\lambda^A\lambda^B + 2 \, {}^{(3)}R_{rB}\lambda^B,
\end{eqnarray}
while the Ricci scalar takes the form
\begin{equation}
	{}^{(3)}R=\frac{2}{\lambda}(\d_rK -\lambda^A\d_AK) + {}^{\gamma}R-
	K^A_BK^B_A-K^2-\frac{2}{\lambda}D_AD^A\lambda.
\end{equation}

The asymptotic conditions considered in section~\ref{sec:hamasymcond} imply
\begin{gather}
	\lambda = 1 + \frac{1}{r} \xbar \lambda +\frac{1}{r^2} \lambda^{(2)}
	+  o(r^{-2}),\quad
	\lambda^A = \frac{1}{r^2} \xbar \lambda^A + \frac{1}{r^3}
	\lambda^{(2)A} + o(r^{-3}),\\
	K^A_B = - \frac{1}{r} \delta^A_B + \frac{1}{r^2} \xbar k^A_B +
	\frac{1}{r^3} {k^{(2)}}^A_B + o(r^{-3}),
\end{gather}
where
\begin{gather}
	\xbar \lambda = \frac{1}{2} \xbar h_{rr}, \quad \xbar \lambda^A = \xbar h_r^A =\xbar
	h_{rB}\xbar \gamma^{BA}, \quad
	\xbar k^A_B = \frac{1}{2} \xbar h^A_B + \xbar \lambda \delta^A_B +
	\frac{1}{2} \xbar D^A \xbar \lambda_B + \frac{1}{2} \xbar D_B \xbar
	\lambda^A.
\end{gather}
The indices on the barred quantities are lowered and raised with $\xbar
\gamma_{AB}$ and its inverse $\xbar \gamma^{AB}$.

\section{More details on the divergences of Lorentz generators}
\label{app:loga}

In this appendix two things are done.

\begin{itemize}
\item First, we give the first two leading orders in the asymptotic expansion of the constraints.
\item  Second, we clarify the appearance (and non-appearance) of logarithmic terms in the asymptotic expansion of the fields. Some explicit examples of solutions to the gravitational constraints equations
with logarithmically divergent Lorentz charges have been 
constructed~\cite{Huang:2010yd}. We make  the link with 
this analysis. 
\end{itemize}
As  before, we  assume $\xbar\lambda^A = 0$.

Expending the momentum constraint $\mathcal H_A$ to second order, we get
\begin{flalign}
	\label{eq:constHAtotsec}
	\mathcal H_A 
	& = -2 \left( \d_r \pi^r_A + \d_B \pi^B_A - {}^\gamma \Gamma_{BA}^C
	\pi^B_C - \frac{1}{\lambda} \d_A \lambda \pi^r_r\right) + o (r^{-1})\\
	& =  -2 \xbar \gamma_{AB} (\xbar \pi^{rB} +\xbar D_C \xbar
	\pi^{BC})\nonumber\\
	& \qquad-
	\frac{2}{r} \left[ \xbar \gamma_{AB} \xbar D_C \pi^{(2)BC} - \d_A
	\xbar \lambda \xbar \pi^{rr} + \xbar D_B (\xbar h_{AC} \xbar \pi^{BC})
	- \frac{1}{2} \xbar \pi^{BC} \xbar D_A \xbar h_{CB}\right] + 
	o(r^{-1}),
\end{flalign}
which, in particular, implies
\begin{equation}
	\label{eq:constHAsub}
\xbar \gamma_{AB} \xbar D_C (\pi^{(2)BC}-2 \xbar
	\lambda \xbar \pi^{BC}) =  \d_A
	\xbar \lambda (\xbar \pi^{rr} - \xbar \pi^B_B)
	- 2\xbar D_B (\xbar k_{AC} \xbar \pi^{BC})
	+ \xbar \pi^{BC} \xbar D_A \xbar k_{CB}.
\end{equation}
If we contract the second line with Killing vectors of the sphere $Y^A$ and
integrate, we obtain three integrability conditions:
\begin{equation}
	\oint d^2x \Big((\xbar \pi^{rr} - \xbar \pi^B_B) Y^A\d_A
	\xbar \lambda  + \xbar \pi^{BC} \cL_Y \xbar k_{BC}\Big) = 0.
\end{equation}
These conditions are necessary and sufficient to guaranty the existence of
$\pi^{(2)AB}$ such that \eqref{eq:constHAsub} is valid. They are the
hamiltonian equivalent of the integrability conditions necessary for the existence of
solutions at second order in the hyperbolic description as described in
\cite{Beig:1983sw,Compere:2011ve} (see also \cite{Beig:1987zz}).

These conditions are consequences of the asymptotic conditions we imposed on
our fields. Looking at \eqref{eq:constHAtotsec}, we see that if the integrability
conditions are not satisfied, they produce a logarithmic term in
$\pi^{rA}$. In other words, if we assume, 
\begin{equation}
	\pi^{rA} = r^{-1} \xbar \pi^{rA} + \log r r^{-2} \pi^{(l)rA} + r^{-2}
	\pi^{(2)rA} + o(r^{-2}),
\end{equation}
then equation \eqref{eq:constHAsub} becomes
\begin{multline}
	\xbar \gamma_{AB} \pi^{(l)rB}=-\xbar \gamma_{AB} \xbar D_C (\pi^{(2)BC}-2 \xbar
	\lambda \xbar \pi^{BC})\\ +  \d_A
	\xbar \lambda (\xbar \pi^{rr} - \xbar \pi^B_B)
	- 2\xbar D_B (\xbar k_{AC} \xbar \pi^{BC})
	+ \xbar \pi^{BC} \xbar D_A \xbar k_{CB}.
\end{multline}
In this case, there is no integrability condition as the logarithmic term will
absorb the corresponding contribution. This logarithmic term will then appear in
the angular momentum charges as a divergent term.

The solution described in \cite{Huang:2010yd} has the following asymptotic
behaviour:
\begin{gather}
	\xbar\pi^{rr} = \sqrt{\xbar \gamma}\Big(\beta(x^A) - B_m Y_{1,m}(x^A)\Big),
	\quad \xbar \pi^{rA} = -\sqrt{\xbar \gamma} \,\xbar D^A\Big(B_m
	Y_{1,m}(x^A)\Big), \\
	\xbar \pi^{AB} = \sqrt{\xbar \gamma} \xbar \gamma^{AB} B_m Y_{1,m}(x^A),\quad 
	\xbar \lambda = \frac{1}{2} A + \frac{1}{4} \alpha(x^A), \quad \xbar
	k_{AB} = A\xbar \gamma_{AB},
\end{gather}
where $A, B_m$ are constants $(m=-1,0,1)$ and $Y_{1,m}$ are $l=1$ spherical
harmonics.
In this case, the integrability condition is
\begin{equation}
	\oint d^2\Omega\, \Big(\beta -3 
	B_m Y_{1,m})\Big)Y^A\d_A \alpha = 0. \label{eq:IntC1}
\end{equation}
The solution given in corollary 3.4 of \cite{Huang:2010yd} corresponds 
to a specific choice of $\beta$ and $\alpha$ for which
this condition is not fulfilled. In that case, the logarithmic term in the
expansion of $\pi^{rA}$ has to be non-zero which will introduce a logarithmic
divergence in the angular momentum charges. One can show that this divergence
will be exactly given by the value of the integrability condition that
reproduces the results of \cite{Huang:2010yd}.

A similar analysis has to be done for the hamiltonian constraint $\mathcal H$.
After some algebra, we get
\begin{flalign}
	\mathcal H & = -\left(1 + 3 \frac{\xbar \lambda}{r}\right)
	 \frac{2}{r^3} \sqrt{\gamma}\left(\xbar D_A \xbar D_B \xbar k^{AB} -
	\xbar D_A \xbar D^A \xbar k\right) \nonumber \\
	& \quad + \frac{1}{r^2} \left\{  - \sqrt{\xbar \gamma} \Big[ \xbar D_A
	\xbar D_B ( h^{(2)AB} + 2 \lambda^{(2)} \xbar \gamma^{AB} + 4\xbar
	\lambda k^{AB})\right. \nonumber\\ &\qquad \qquad \qquad \qquad  - \xbar D_A
	\xbar D^A ( h^{(2)} + 4 \lambda^{(2)}  + 4\xbar
	\lambda k) - ( h^{(2)} + 4 \lambda^{(2)} + 4\xbar
	\lambda k)\Big] \nonumber \\ 
	&\qquad \qquad - \sqrt{\xbar \gamma} \Big[ 3 \xbar k^A_B \xbar k^B_A + 4 \xbar k^{AB} \xbar D_C \xbar D^C
	\xbar k_{AB} + 4 \xbar k^{AB} \xbar D_A \xbar D_B \xbar k - 4 \xbar k^{AB} ( \xbar D_C \xbar D_A
	+ \xbar D_A \xbar D_C) \xbar k^C_A 
	\nonumber \\ & \qquad \qquad \qquad \qquad  + 3
	\xbar D_A \xbar k_{BC} \xbar D^A \xbar k^{BC}- \xbar D_A k \xbar D^A k - 2 \xbar D_A
	\xbar k_{BC} \xbar D^B \xbar k^{AC}
	\nonumber \\ & \qquad \qquad \qquad \qquad + 4 \xbar D_B
	\xbar k^B _A ( \xbar D^A \xbar k - \xbar D_C \xbar k^{CA} ) - \xbar
	k^2+12 \xbar \lambda^2 + 6 \xbar
	\lambda \xbar D_C \xbar D^C \xbar \lambda + 4 \xbar D_A \xbar \lambda
	\xbar D^A \xbar \lambda\Big]\nonumber \\ & \left.\qquad \qquad + 
	\frac{1}{\sqrt{\xbar\gamma}} \Big[ 2 \xbar \pi^{rA} \xbar \pi^r_A +
	\xbar \pi^{AB} \xbar \pi_{AB} + \frac{1}{2} (\xbar\pi^{rr} - \xbar\pi^A_A)^2 - (\xbar
	\pi^A_A)^2\Big]
	\right\} + o(r^{-2})
\end{flalign}
The sub-leading contribution takes the form
\begin{equation}
	\label{eq:constHsecond}
	\sqrt{\xbar \gamma} \left(\xbar D_A \xbar D_B \alpha^{AB} - 
	\xbar D_A \xbar D^A \alpha - \alpha\right) = J
\end{equation}
where $\alpha^{AB}$ is linear in the second-order perturbations of the fields, and $J$ quadratic in their first-order perturbations.  Integrating this equation with a "boost" parameter $b$ such that $\xbar D_A \xbar D_B b
+ \xbar \gamma_{AB} b = 0$, we get the three integrability conditions:
\begin{equation}
	\oint d^2x \, b J = 0. \label{eq:IntC2}
\end{equation}
on the first-order perturbation.
As before, one can show that if $J$ satisfies these identities, then there
exists a $h^{(2)}$ such that the constraint is valid. If they do not hold, then
logarithmic terms will appear in the expansion of $g_{ij}$.

The same analysis on the radial constraint $\cH_r$ does not give any new
integrability conditions. It takes the form
\begin{multline}
	\cH_r = - \frac{2}{r} \left( \d_A \xbar \pi^{Ar} - \xbar
	\pi^A_A\right) - \frac{2}{r^2} \Big( -\pi^{(2)rr} -
	{\pi^{(2)}}^A_A\\ - \xbar \lambda (\xbar \pi^{rr} - \xbar\pi^A_A) -
	\xbar k_{AB} \xbar\pi^{AB} + \d_A (\pi^{(2)Ar} + 2 \xbar\lambda 
	\xbar \pi^{Ar})\Big)+ O(r^{-3}),
\end{multline}
which imposes
\begin{equation}
	\xbar\pi^A_A =  \d_A \xbar \pi^{Ar} \qquad \Rightarrow \qquad \xbar
	D_A\xbar D_B \xbar \pi^{AB} + \xbar \pi^A_A = 0,
\end{equation}
and fixes $\pi^{(2)rr}-\pi^{(2)A}_{\phantom{(2)A}A}$ in terms of the 
other asymptotic fields.

The six integrability conditions (\ref{eq:IntC1}), (\ref{eq:IntC2}) for the existence of the subleading terms without the need to introduce logarithms are easily verified to be fulfilled by
\begin{itemize}
	\item  spacetimes satisfying the R-T parity conditions~\cite{Regge:1974zd}:
		\begin{equation}
			\xbar \lambda \sim \xbar \pi^{rA} \sim \xbar k_{AB}\sim \text{even }, \qquad
			\xbar \pi^{rr} \sim \xbar \pi^{AB} =
			\text{odd},
		\end{equation}
	\item  spacetimes satisfying the new parity conditions:
		\begin{equation}
			\xbar \lambda \sim \xbar \pi^{AB} \sim \text{even }, \qquad
			(\xbar \pi^{rr} - \xbar \pi^A_A) \sim \xbar k_{AB}\sim \xbar \pi^{rA} =
			\text{odd}.
		\end{equation}
\end{itemize}
The parity conditions guarantee therefore the consistency of the perturbative expansion adopted here.

\section{BMS algebra in the ``instant form'' and in the hyperbolic form of the dynamics}
\label{app:Hyperbolic}

We relate in this appendix the results of Section \ref{sec:BMSAlgebra}
to the recent work on the asymptotic symmetry algebra carried out in \cite{Troessaert:2017jcm} in the context of the hyperbolic treatment of spatial infinity \cite{Ashtekar:1978zz,BeigSchmidt,Beig:1983sw}, with a Hamiltonian that generates boosts asymptotically.  The results of \cite{Troessaert:2017jcm} apply to the analysis of 
\cite{Compere:2011ve}, which adopts different boundary conditions than the
ones taken here (logarithmic terms, no parity condition) and regularizes the
resulting infinities in the framework of holographic renormalization.
Nevertheless, there is an overlap in the corresponding symmetries.  

As described in \cite{Troessaert:2017jcm}, the algebra of the asymptotic symmetries  is the semi-direct
sum of the Killing vectors of the unit hyperboloid with a set of 
super-translations. If we use the following metric on the hyperboloid
\begin{equation}
	h^{0}_{ab} dx^a dx^b = - \frac{1}{(1-s^2)^2} ds^2 + \frac{1}{(1-s^2)}
	\xbar \gamma_{AB} dx^Adx^B,
\end{equation}
then the Lorentz algebra is generated  by
\begin{equation}
	\label{eq:hyperbollorentz}
	\mathcal Y^s = - (1-s^2) b, \quad \mathcal Y^A = Y^A  -
	s \xbar D^A b,
\end{equation}
and the  abelian algebra of super-translations is parametrized by functions
$\omega$ on the
hyperboloid satisfying
\begin{equation}
	\label{eq:hyperbolsuper}
	(\mathcal D_a \mathcal D^a + 3)\, \omega = -(1-s^2)^2 \d_s^2 \omega 
	+(1-s^2) \xbar D_A \xbar D^A \omega + 3 \omega= 0,
\end{equation}
where $\mathcal D_a$ is the covariant derivative associated to the metric $h^0_{ab}$.
As shown in \cite{Troessaert:2017jcm}, in the limit $s \rightarrow \pm 1$, the two branches of
solutions to the super-translation equation \eqref{eq:hyperbolsuper} have a different
behaviour. The one corresponding to odd functions $\hat\omega = \sqrt{1-s^2}\,
\omega$ under the combined
action of a time reversal and antipodal map tends to finite functions on the
sphere and describes the usual BMS
super-translations. The algebra of asymptotic symmetries parametrized by
\eqref{eq:hyperbollorentz} and \eqref{eq:hyperbolsuper} with odd $\hat\omega$'s is then the usual BMS asymptotic symmetry algebra at null infinity.

More explicitly, the BMS supertranslation parameter $\mathcal T(x^A)$ can be
obtained through the following construction. An odd function $\omega = (1-s^2)^{-\frac 1 2} \hat \omega$ 
solution to
equation \eqref{eq:hyperbolsuper} has a spherical harmonics
expansion given by
\begin{equation}
	\label{eq:gensolomega}
	\hat \omega = \sum_{lm} \omega_{l,m} \Psi_l(s) Y_{l,m}(x^A), 
	\quad \Psi_l =\frac 1 2  (1-s^2)^2 \d_s^2Q_l.
\end{equation}
The functions $Q_l(s)$ are Legendre
functions of the second kind and can be written in terms of Legendre
polynomials $P_l(s)$ as
\begin{equation}
	Q_l(s) = P_l(s) \frac 1 2 \log \left (\frac{1+s}{1-s}\right) + \tilde
	Q_l(s),
\end{equation}
where $\tilde Q_l(s)$ are polynomials. The action of the Lorentz algebra on $\hat\omega$ is given by
\begin{equation}
	\label{eq:actionlorentzomega}
	\delta_{Y,b} \hat \omega = Y^A \d_A \hat \omega - sb \hat\omega -s \xbar D^A b \d_A \hat \omega - (1-s^2)
	b \d_s \hat \omega.
\end{equation}
Defining the  BMS supertranslation parameter $\mathcal T(x^A) = \lim_{s\to 1}
\hat\omega$, we can
evaluate the above identity at $s=1$ using the asymptotic behaviour $\psi_l(s) = 1
+ O(1-s)$ to obtain
\begin{equation}
	\label{eq:usualBMS4super}
	\delta_{Y,b} 
	\mathcal T = Y^A \d_A
	\mathcal T - b \mathcal T - \xbar D^Ab \d_a \mathcal T,
\end{equation}
This is the usual action of a Lorentz transformation on BMS supertranslations.

The difference between the description of the BMS algebra given in
\eqref{eq:hamilbmsI}-\eqref{eq:hamilbmsIV} and the description given in
equation \eqref{eq:usualBMS4super} is in the choice
of representative functions used to
parametrize the super-translations $\omega$. In order to recover the ADM
description of the supertranslation, we have to define $W$ and $T$ as initial
conditions at $s=0$:
\begin{equation}
	\label{eq:initcondomega}
	\omega\vert_{s=0} = \hat\omega\vert_{s = 0} = W(x^A), \quad 
	\d_s\omega\vert_{s=0} = \d_s \hat \omega\vert_{s=0} = T(x^A).
\end{equation}
We see that the hyperboloid function $\hat \omega$ is odd if and only if $W$
and $T$ are respectively an odd and an even function on the sphere. From the
action of Lorentz algebra on $\hat\omega$ given in equation
\eqref{eq:actionlorentzomega}, we can derive the corresponding action on $W$
and $T$:
\begin{flalign}
	\delta_{Y,b} T & = Y^A \d_A T - \xbar D^A b \d_A W - b \xbar D_A \xbar
	D^A W - 3 b W, \\
	\delta_{Y,b} W & = Y^A \d_A W - bT.
\end{flalign}

The change of basis from the pair of functions $(W,T)$ to the BMS
supertranslation parameter
$\mathcal T$ is obtained by solving equation \eqref{eq:hyperbolsuper} with the
initial condition \eqref{eq:initcondomega} and then defining
\begin{equation}
	\mathcal T = \lim_{s \to 1} \Big( \sqrt{1-s^2} \, \omega \Big).
\end{equation}
Expending all quantities in spherical harmonics and using the general solution
written in \eqref{eq:gensolomega}, we can write this change of
basis explicitly
\begin{gather}
	\mathcal T = \sum_{lm} \omega_{l,m} Y_{l,m} (x^A), \\
	W = \sum_k \sum_{m=-2k-1}^{2k+1} W_{2k+1,m}\, Y_{2k+1,m}, \quad
	T = \sum_k \sum_{m = -2k}^{2k} T_{2k,m} \, Y_{2k,m},\\
	\omega_{2k+1,m}\, \psi_{2k+1}\vert_{s=0} =W_{2k+1,m}, \quad
	\omega_{2k,m} \, \d_s\psi_{2k}\vert_{s=0} = T_{2k,m}.
\end{gather}
The first few $\psi_l$ functions can be easily computed
\begin{equation}
	\psi_0 = s, \quad \psi_1 = 1, \quad \psi_2 = \frac 3 4 (1-s^2)^2\log
	\left (\frac{1+s}{1-s}\right) + \frac 5 2 s (1-s^2),
\end{equation}
and we can use them to write the first few component of the change of basis:
\begin{equation}
	\omega_{0,0} = T_{0,0}, \quad \omega_{1,m} = W_{1,m}, \quad
	\omega_{2,m} = \frac 1 4 T_{2,m}.
\end{equation}

\end{appendix}

\end{document}